\documentclass[10pt]{article}
\def\be{\begin{equation}}
\def\ee{\end{equation}}
\def\bea{\begin{eqnarray}}
\def\eea{\end{eqnarray}}

\begin{document}
\thispagestyle{empty}
\begin{titlepage}
\begin{center}
\vskip 1.5cm
{\LARGE \bf Gravitational perturbations of the Kerr black hole}
\vskip 0.5cm
{\LARGE \bf due to arbitrary sources}
\end{center}
\normalsize
\vskip1cm
\begin{center}
{\large \bf Claudia Moreno}
\footnote{E-mail: yaya@fis.cinvestav.mx}
{\large and} {\large \bf Dar\'{\i}o N\'{u}\~{n}ez}
\footnote{E-mail: nunez@nuclecu.unam.mx}
\end{center}
\begin{center}
\baselineskip=13pt {\large \it Center for Gravitational Physics \& Geometry \\
Penn State University, University Park, PA 16802}
\vskip0.5cm
{\large \it and}
\vskip0.5cm
$^1${\large \it Departamento de F{\'\i}sica,\\
Centro de Investigaci{\'o}n y de Estudios Avanzados del I. P. N.,\\
A. P. 14-740, 07000 M{\'e}xico, D.F., MEXICO}

$^2${\large \it Instituto de Ciencias Nucleares - UNAM,\\
Apdo. 70-543, Ciudad Universitaria, 04510 Mexico, D.F.}
\vglue 0.8cm
\end{center}
\date{\today}

\begin{abstract}
We describe the Kerr black hole in the ingoing and outgoing
Kerr-Schild horizon penetrating coordinates. Starting from the
null vector naturally defined in these coordinates, we construct
the null tetrad for each case, as well as the corresponding
geometrical quantities allowing us to explicitly derive the field
equations for the ${\Psi_0}^{(1)}$ and ${\Psi_4}^{(1)}$ perturbed
scalar projections of the Weyl tensor, including arbitrary source
terms. This perturbative description, including arbitrary sources,
described in horizon penetrating coordinates is desirable in
several lines of research on black holes, and contributes to the
implementation of a formalism aimed to study the evolution of the
space time in the region where two black holes are close.
\end{abstract}

\vfill

\end{titlepage}

\section{Introduction}

A particle falling unto a black hole is an efficient mechanism for generation of
gravitational waves. This problem is not only of great importance to gravitatonal
wave astrophysics, but it also has important application of numerical general
relativity.

The fact that several gravitational observatories (LIGO, VIRGO, GEO600, and later
LISA)will start to be working soon is one of the most exciting news lately received
by the general relativistic community. Those observatories will allow a much deeper
insight on the phenomena near the black holes such as the above mentioned particle
falling unto it, and later, will allow us to see beyond the recombination era, plus
all the not yet expected ones.

These technological achievements, together with a strong development in the
computational techniques, definitely boost the great theoretical knowledge acquired
in the last century. In particular, the theoretical study on gravitational
perturbations goes back to Einstein and has produce excellent works such as the one
by Chadrasekhar \cite{libro} or the one by Frolov \cite{frol}, for just to mention a
couple of them. Also, there is the work by Teukolsky \cite{Teukolsky} who, using the
Newman Penrose formalism for null tetrads (see \cite{libro,frol}), was able to obtain
an  equation for the ingoing and outgoing radiative parts of the perturbed Weyl
tensor, namely ${\Psi_0}^{(1)}$ and ${\Psi_4}^{(1)}$, respectively, which were
separable in the frequency domain. The radial part of those equation were useful in
obtaining information about the asymptotic behavior of the gravitational
perturbations in the vacuum case.

The complexity of the equations prevented further advances until the development of
numerical techniques for studding the evolution of the equations. There was large
effort in understanding the collision of black holes evolving directly the Einstein
equations \cite{Miguel,Pablo1}, as well as those studies comprising semi analytical
results where they used the perturbation equations derived by Teukolsky with two
important changes: First, with a coordinate system which avoids the coordinate
singularity, and second, they worked in the time domain, to be able to study the
evolution.

Campanelli et al. \cite{Pablo2} presented a work in this last direction and obtained
the evolution of the ${\Psi_4}^{(1)}$ Weyl tensor for the case without sources. In
the present work, we continue within such line of research, obtaining the equations
which allow the study by numerical evolution of the ${\Psi_0}^{(1)}$ and
${\Psi_4}^{(1)}$ perturbed Weyl scalars in the case with arbitrary sources generating
the perturbation. Furthermore, families of initial data with appealing features such
as describing regularly multiple moving spinning black holes \cite{Dier}, use the Kerr
Schild horizon penetrating coordinates.

We stress the fact that in the present work we explicitly derive the field equations
for the ${\Psi_0}^{(1)}$ and ${\Psi_4}^{(1)}$ perturbed scalar projections of the Weyl
tensor, including arbitrary source terms. Even though there is the coordinate
transformation from the Boyer Lindquist  to the Kerr Schild ones, the change of
tetrad involves rotations and dilatation of the null tetrad which prevent to treat the
Teukolsky equation as a tensor operator, and it is necessary to derive the equations
from their general definitions, as is done in the present work.

The work is composed as follows. In the next section we derive the
Kerr metric in ingoing and outgoing Kerr Schild coordinates, and
present the field equations for ${\Psi_0}^{(1)}$ and
${\Psi_4}^{(1)}$, deriving the master field equation. Next, we
take the naturally defined null vector and construct the null
tetrad, the differential operators, the spinor coefficients and
the Weyl scalars and, describing the main steps in the deduction,
we finally derive explicitly the expression for the perturbation
equations for arbitrary sources valid in a Kerr Schild
description, in the ingoing as well as in the outgoing
formulation. We test our result in simple examples and conclude
mentioning several case where will be of interest apply the
obtained equations.

\section{Kerr Schild line element and master perturbation equation} \label{s:KS}

The rotating black hole is usually described by the line element in
Boyer Lindquist coordinates \cite{MTW}:
\begin{equation}
ds^{2}=\frac{\Delta }{\Sigma }\left( dt-a\sin ^{2}\theta d\varphi \right)
^{2}-\frac{\sin ^{2}\theta }{\Sigma }\left[ \left( r^{2}+a^{2}\right)
d\varphi -adt\right] ^{2}-\Sigma \left( \frac{dr^{2}}{\Delta }
+d\theta^{2}\right),
\end{equation}
with $\Delta =r^{2}-2Mr+a^{2}$, and $\Sigma =r^{2}+a^{2}\cos ^{2}\theta $.

However, as mentioned above, this description has a divergence at
the external horizon: $r_+=M(1+\sqrt{1+({{a}\over{M}})^2})$, which
prevent that the evolution analysis could be performed in the
vicinity of the external horizon. Moreover, it has been
established that the  gravitational wave emission of several
phenomena such as the matter accretion or the inspariling of an
object unto the black hole, occurs precisely within this region
\cite{Thorne}. Detailed studies show that there is a sweep upward
in frequency, during the last 15 minutes before the collision and
final coalescence, that is, precisely in the region close to the
external horizon for which the Boyer Lindquist coordinates
are not well suited.

Thus it is imperative to be able to perform the description of the
phenomena within a coordinate system which does not have those
coordinate problems. Whit in this spirit the Kerr Shild
description not only avoids the problem of coordinate
singularities at the horizon but is the most natural description
to be matched with the null tetrad formulation. In this way,
following \cite{libro}, we define the coordinates $t$ and
$\varphi$ in terms of the outgoing and ingoing ones, $u_{{\rm
in}\atop{\rm out}}$:

\begin{equation}
du_{{\rm in}\atop{\rm out}}=dt\pm \frac{r^{2}+a^{2}}{\Delta }dr,
\qquad \qquad d\varphi_{{\rm in}\atop{\rm out}}
=d\varphi \pm \frac{a}{\Delta }dr,  \label{uno}
\end{equation}
and, defining in each case a new temporal coordinate

\be
dt_{{\rm in}\atop{\rm out}}=du_{{\rm in}\atop{\rm out}}\mp dr,
\ee
We obtain that the Kerr space time is described as:
\begin{eqnarray}
ds^{2} &=&(1-\frac{2Mr}{\Sigma}){dt^2}_{{\rm in}\atop{\rm out}}
\mp \frac{4Mr}{\Sigma}dt_{{\rm in}\atop{\rm out}} dr +
\frac{4Mra\sin^2\theta}{\Sigma }dt_{{\rm in}\atop{\rm out}}
d\varphi_{{\rm in}\atop{\rm out}}-
\left(1+\frac{2Mr}{\Sigma}\right)dr^2 \nonumber \\ && \pm
2\left(1+\frac{2Mr}{\Sigma}\right)a\sin^2\theta dr d\varphi_{{\rm
in}\atop{\rm out}} -\Sigma d\theta^2-\left(
r^2+a^2+\frac{2Mra^2\sin^2\theta}{\Sigma}\right )\sin^2\theta
d{\varphi^2}_{{\rm in}\atop{\rm out}}, \label{eq:ks}
\end{eqnarray}
which is the known Kerr Schild description. Noticing that this last expression for
the Kerr line element, Eq. (\ref{eq:ks}) can be rewritten as:
\be
ds^2=({\eta_{\mu\nu}}_{{\rm in}\atop{\rm out}}-{{2Mr}\over{\Sigma}}{l_\mu}_{{\rm
in}\atop{\rm out}}\,{l_\nu}_{{\rm in}\atop{\rm out}})dx^\mu dx^\nu, \ee with
${\eta_{\mu\nu}}_{{\rm in}\atop{\rm out}}$ is the flat space metric, which in this
case takes the form: \be {\eta_{\mu\nu}}_{{\rm in}\atop{\rm out}}dx^\mu
dx^\nu={dt^2}_{{\rm in}\atop{\rm out}}-dr^2-\Sigma d\theta^2-(r^2+a^2)\sin^2\theta
d{\varphi^2}_{{\rm in}\atop{\rm out}}\pm2\,a\sin^2\theta dr d\varphi_{{\rm
in}\atop{\rm out}},
\ee and ${l_\mu}_{{\rm in}\atop{\rm out}}$ is a null vector, null
as much for the flat metric as for the complete one, and is given by:
\be
{l_\mu}_{{\rm in}\atop{\rm out}}=(1,\pm 1,0,-a\sin^2\theta). \label{eq:l}
\ee

We want to stress the fact that, as mentioned above, the line element in the Kerr
Schild coordinates, Eq.(\ref{eq:ks}), not only has no geometric divergence all the
way from infinity to the interior of the black hole, but also defines naturally one
of the null vectors.

Thus, we use this null vector, defined by Eq.(\ref{eq:l}), to construct the null
tetrad, the differential operators, the spinor coefficients and the Weyl scalars,
which are needed to study the gravitational perturbations within the Newman Penrose
formulation. The null tetrad $l, n, m $ and $m^*$ is defined by the following
identities:
\begin{equation}
l\cdot n=1\qquad {\rm and}\qquad m \cdot m^*=-1, \label{eq:met1}
\end{equation}
and from the relation of the tetrad with the metric coefficients: \be
g_{\mu\nu}=2[l_{(\mu}n_{\nu)}-m_{(\mu}{m^*}_{\nu)}]. \label{eq:met2}
\ee

Before presenting explicitly the ingoing and outgoing cases, we
derive the field equations for the perturbed Weyl components,
valid for any algebraically special type D space time.

Perturbing the Bianchi identities in the NP formulation \cite{libro,frol}, taking
into account the expressions derived in the above section, coupled equations for the
perturbed Weyl scalars ${\Psi_0}^{(1)}$ and ${\Psi_1}^{(1)}$ and ${\Psi_3}^{(1)}$
${\Psi_4}^{(1)}$, are obtained. Using the generalized commutation relations between
the spinor operators given in the appendix A, Eq.(\ref{a:con}), as well as the
perturbed Ricci identities obtained form \cite{libro,frol} and the expressions
derived in the above section, the following separate equations for the perturbed Weyl
scalars ${\Psi_0}^{(1)}$ and ${\Psi_4}^{(1)}$ are obtain: \bea &&{\lbrace}({\bf{D}}
-4\rho -\rho^*-3\varepsilon +\varepsilon^* )({\bf{\Delta}}+\mu-4\gamma )-
({\bf{\delta}}-2\beta-4\tau)({\bf{\delta}}^*+4\beta^*-3\pi)-3\Psi_2 {\rbrace}{\Psi_0}^{(1)} \nonumber \\
&&= 8\pi\,{}_2{\cal{T}}{^{\mu\nu}}\,{T_{\mu\nu}}^{(1)},
\label{eq:teu1} \\
&&{\lbrace}({\bf{\Delta}}+4\mu+\mu^*+3\gamma -\gamma^* )({\bf{D}}-\rho+4\varepsilon )-
({\bf{\delta}}^*-2\beta^*+7\pi-\tau^* )({\bf{\delta}}+4\beta-\tau )-3\Psi_2
{\rbrace}\,{\Psi_4}^{(1)} \nonumber \\
&&=8\pi\,{}_{-2}{\cal{T}}{^{\mu\nu}}\,{T_{\mu\nu}}^{(1)}, \label{eq:teu2} \eea where
${}_s{\cal{T}}^{\mu\nu}$ are given by: \bea &&
{}_2{\cal{T}}^{\mu\nu}=-({\bf{\delta}}-2\beta-4\tau)({\bf{\delta}}-\pi^*)\,l^ \mu\,l^
\nu
-({\bf{D}}-4\rho-\rho^*-3\varepsilon+\varepsilon^*)({\bf{D}}-\rho^*-
2\varepsilon+2\varepsilon^*)\,m^\mu\,m^\nu \nonumber \\
&&+[({\bf{\delta}}-2\beta-4\tau)({\bf{D}}-2\rho^*-2\varepsilon)+({\bf{D}}-4\rho-\rho^*-3\varepsilon+
\varepsilon^*)({\bf{\delta}}-2\beta+2\pi^*)]\,l^{(\mu}\,m^{\nu)}, \nonumber \\
&& {}_{-2}{\cal{T}}^{\mu\nu}= \nonumber \\
&& -({\bf{\delta}}^*-2\beta^*+7\pi-\tau^*)({\bf{\delta}}^*+2\pi-\tau^*)\,n^\mu\,n^\nu
-({\bf{\Delta}}+4\mu+\mu^*+3\gamma-\gamma^*)({\bf{\Delta}}+\mu^*+2\gamma-2\gamma^*)\,{m^*}^\mu\,{m^*}^\nu \nonumber \\
&&+[({\bf{\delta}}^*-2\beta^*+7\pi-\tau^*)({\bf{\Delta}}+2\mu^*+2\gamma)+({\bf{\Delta}}+4\mu+\mu^*+
3\gamma-\gamma^*)({\bf{\delta}}^*-2\beta^*+2\pi-2\tau^*)]\,n^{(\mu}\,{m^*}^{\nu)}, \nonumber \\
\eea with ${T_{\mu\nu}}^{(1)}$ is the source of the perturbation. These equations
were first derived by Teukolsky in 1972, \cite{Teukolsky} and, as long as he was
interested in the asymptotic behavior, the Eddington Finkelstein coordinates which he
used, were well suited for the analysis.

The studies are performed on the ${\Psi_0}^{(1)}$ and ${\Psi_4}^{(1)}$ perturbed Weyl
scalars as long as those are the most significant components in the gravitational
radiation \cite{Teukolsky}.

Acting on $\rho^{-4}\,{\Psi_4}^{(1)}$ and using the commutation
relations between the operators, Eq.(\ref{a:con}), and the action
of the operators on the spinor coefficients, Eq.(\ref{a:spin})
into the appendix A, the perturbation equations for
${\Psi_0}^{(1)}$ and $\rho^{-4}\,{\Psi_4}^{(1)}$,
Eqs.(\ref{eq:teu2}), can be written as a single master equation in
terms of the parameter $s$: \bea && {\lbrace}{\bf \Delta}{\bf
D}-{\bf \delta}^*{\bf \delta}
+(\mu^*-(2s+1)\,\gamma-\gamma^*)\,{\bf D}-
((2s+1)\rho+2s\varepsilon)\,{\bf \Delta}-(2(s+1)\beta^*-(2s+1)\pi-\tau^*)\,{\bf \delta}\nonumber \\
&& +(2s\beta+(2s+1)\tau)\,{\bf \delta}^*+s\,(2s+1)[2(\rho+\varepsilon)\,\gamma-2\beta\,\pi-2\pi\,\tau+
2\beta^*\,\tau-\Psi_2]- \nonumber \\
&&2\,s\,[({\bf \Delta}\,\varepsilon)-({\bf
\delta^*}\beta)+\varepsilon\,(\mu^*-\gamma^*)+\beta\,\tau^*]+
4\,s\,(s+1)\,\beta\,\beta^*{\rbrace}{}_s\,\psi=8\pi\,\rho^{(s-2)}\,{}_{s}{\cal{T}}{^{\mu\nu}}\,{T_{\mu\nu}}^{(1)},
\label{eq:Tf}
\eea
with ${}_2\,\psi ={\Psi_0}^{(1)}$ y ${}_{-2}\,\psi
=\rho^{-4}\,{\Psi_4}^{(1)}$. Even though the ${\Psi_0}^{(1)}$ and ${\Psi_4}^{(1)}$
perturbed Weyl scalars are the most significant components in the gravitational
radiation, it would be interesting to analyze the evolution of the rest of the
perturbed Weyl scalars, ${\Psi_1}^{(1)}$, ${\Psi_2}^{(1)}$ and ${\Psi_3}^{(1)}$. Some
studies indicate that it is possible to obtain a single master equation for the five
of the perturbed Weyl scalars \cite{yo,Chrz,PhD}. These studies are currently under
investigation and will be reported elsewhere.

\section{Null tetrad formulation and explicit derivation of the perturbation equations
for ${\Psi_0}^{(1)}$ and ${\Psi_4}^{(1)}$ with sources}

Next, we presents the main steps in the derivation of the explicit
form of the field equations for the perturbed ${\Psi_0}^{(1)}$ and
${\Psi_4}^{(1)}$ Weyl components with sources. In the appendix B,
we present the main expressions for the action of the operators on
the spinor coefficients, needed in the derivation of the explicit
equation. We are going to present both cases, the ingoing and the
outgoing.

\subsection{Ingoing case} \label{s:pert}

Starting form the naturally defined null vector, given by
Eq.(\ref{eq:l}), using the equations
Eqs.(\ref{eq:met1},\ref{eq:met2}), we construct the following
ingoing null tetrad:
\begin{eqnarray}
{l_{\mu}}_{\rm in} &=&\left( 1,1,0,-a\sin^{2}\theta \right), \nonumber \\
{n_{\mu}}_{\rm in} &=&\frac{1}{2\,\Sigma}\left(\Delta, r^2+a^2(\cos^{2}\theta-\sin^{2}\theta)+
2Mr,0,-a\Delta\sin^{2}\theta \right), \nonumber \\
{m_{\mu}}_{\rm in}
&=&\frac{1}{\sqrt{2}\,(r+i\,a\,\cos\theta)}\left[-i\,a\,\sin\theta,-i\,a\,\sin\theta,
\Sigma,i\,(r^2+a^2)\,\sin\theta \right].
\end{eqnarray}

\noindent From which we obtain the covariant null vectors:
\begin{eqnarray}
{l^{\mu}}_{\rm in} &=&\left( 1,-1,0,0 \right), \nonumber \\
{n^{\mu}}_{\rm in} &=&\frac{1}{2\,\Sigma}\left( r^2+a^2+2Mr, \Delta ,0,2a \right) ,\nonumber  \\
{m^{\mu}}_{\rm in}
&=&-\frac{1}{\sqrt{2}\,(r+i\,a\,\cos\theta)}\left(ia\sin
\theta,0,1,i\csc \theta\right), \label{eq:nulltin}
\end{eqnarray}

\noindent as well as the corresponding differential operators:

\begin{eqnarray}
\bf{D}_{\rm in}&=&{l^{\mu}}_{\rm in}\partial_{\mu }=\partial_{t_{\rm in}} -\partial_r,\nonumber \\
\bf{\Delta}_{\rm in} &=&{n^{\mu}}_{\rm in}\partial_{\mu
}=\frac{1}{2\,\Sigma}\left[(r^2+a^2+2Mr)\partial_{t_{\rm in}}+
\Delta\partial_r+2a\partial_{\varphi_{\rm in}} \right], \nonumber\\
\bf{\delta}_{\rm in}&=&{m^{\mu }}_{\rm in}\partial_{\mu
}=-\frac{\rho_{\rm in}}{\sqrt{2}}(ia\sin
\theta\partial_{t_{\rm in}} +\partial_\theta+i\csc
\theta\partial_{\varphi_{\rm in}}), \label{op:in}
\end{eqnarray}
with the spinor coefficient $\rho_{in}$ defined below in Eq.(\ref{eq:espin}).

The spinor coefficients are computed directly from their definition
\cite{libro,frol}, obtaining:
\begin{eqnarray}
&& \kappa_{\rm in}=\sigma_{\rm in}=\lambda_{\rm in}=\nu_{\rm in}=0, \nonumber \\
&& \rho_{\rm in}=\frac{1}{r+i\,a\,\cos\theta},\quad
2\varepsilon_{\rm in}=\rho_{\rm in }-\rho^*_{\rm in}, \quad
\mu_{\rm in} =\frac{\Delta}{2\,\Sigma}\,\rho_{\rm in},\quad 2
\gamma_{\rm in} = \mu_{\rm in}+
{\mu^*}_{\rm in}-{{(r-M)}\over{\Sigma}}, \nonumber \\
&&\pi_{\rm in} =\frac{ia\sin \theta}{\sqrt{2}\Sigma},\quad 2
\beta_{\rm in}=-{{\cot\theta}\over{\sqrt{2}}}\,\rho_{\rm
in}+\pi^*_{\rm in}+\tau_{\rm in},\quad \alpha_{\rm in} =\pi_{\rm
in}-{\beta_{\rm in}}^*, \quad \tau_{\rm in} =-\frac{ia\sin
\theta}{\sqrt{2}}\,{\rho_{\rm in}}^2,  \label{eq:espin}
\end{eqnarray}
and finally, the only non zero Weyl scalar is:
\begin{equation}
{\Psi _{2}}_{\rm in}=-M{\rho_{\rm in}}^3. \label{eq:esp}
\end{equation}

Using the explicit form of the differential operators,
Eqs.(\ref{op:in}), of the spinor coefficients and the Weyl scalar
Eqs. (\ref{eq:espin}, \ref{eq:esp}), for the null tetrad
Eq.(\ref{eq:nulltin}) of the Kerr metric described in Kerr Schild
coordinates, Eq.(\ref{eq:ks}), with the help of the expressions
for the action of the operators on the spinor coefficients given
in the appendix A, Eq.(\ref{a:spin}), (\ref{a:spinII}), we derive
the equations for the perturbations of the ${\Psi_0}^{(1)}$ and
${\Psi_4}^{(1)}$ perturbed Weyl scalars due to arbitrary source
terms in ingoing coordinates: \bea &&\lbrack
(\Sigma+2\,M\,r){{\partial^2}\over {\partial t^2_{\rm in}}}-
4\,M\,r\,{{\partial^2}\over{\partial r
\partial t_{\rm in} }} -\Delta\,{{\partial^2}\over{\partial r^2}}
-2\,a\,{{\partial^2}\over{\partial r\partial \varphi_{\rm in} }}
-{{\partial^2}\over{\partial
\theta^2}}-{{1}\over{\sin^2\theta}}{{\partial^2}\over{\partial
\varphi^2_{\rm in}}} \nonumber \\
&& - 2\left((s+1)(r-M)+2s\Delta\,\varepsilon_{\rm
in}\right){{\partial}\over{\partial r}}
-2\left(s\,r+(s+1)M-s\,\varepsilon_{\rm
in}\,(\Sigma+4\,M\,r)\right){{\partial}\over{\partial t_{\rm
in}}}\nonumber \\
&&-\left(\cot\theta+4s\,r\tan\theta\,\varepsilon_{\rm
in}\right){{\partial}\over{\partial \theta}}
-2s\left(i{{\cos\theta}\over{\sin^2\theta}}+4\,a\,\varepsilon_{\rm
in}\right){{\partial}\over{\partial \varphi_{\rm in}}}
-s(1-s\cot^2\theta)+{{16a^2}\over{\Sigma}}(1-{{2Mr\cos^2\theta}\over{\Sigma}})
\nonumber \\
&&-4s\varepsilon_{\rm
in}\,\left((s+1)r+(r^2-s(r^2-a^2\cos^2\theta)){{M}\over{\Sigma}}\right)\rbrack{}_s\psi=
16\pi\,\Sigma\,{}_{s}{\cal{T}}{^{\mu\nu}}\,{T_{\mu\nu}}^{(1)},
\label{eq:Tfin}\eea

\noindent where ${}_{s}{\cal{T}}{^{\mu\nu}}$ are given by the
following expressions:

\bea
{}_{2}{\cal{T}}{^{\mu\nu}}&=&{}_{2}{\cal{T}}^{ll}l^\mu\,l^\nu+{}_{2}{\cal{T}}^{mm}m^\mu\,m^\nu+
{}_{2}{\cal{T}}^{lm}l^\mu\,m^\nu,
\nonumber \\
{}_{-2}{\cal{T}}{^{\mu\nu}}&=&{}_{-2}{\cal{T}}^{nn}n^\mu\,n^\nu+{}_{-2}{\cal{T}}^{m^*m^*}{m^*}^\mu\,{m^*}^\nu+
{}_{-2}{\cal{T}}^{nm^*}n^\mu\,{m^*}^\nu,  \label{eq:tau1}
\eea with

\bea {}_{2}{\cal{T}}^{ll}&=&
 {{\rho^2_{\rm in}}\over{2}}\lbrace a^2\sin^2\theta{{\partial^2}\over{\partial
t^2_{\rm in} }} -2ia\sin\theta{{\partial^2}\over{\partial\theta
\partial t_{\rm in} }} +2a{{\partial^2}\over{\partial \varphi_{\rm
in} \partial t_{\rm in} }}
+\csc^2\theta{{\partial^2}\over{\partial \varphi^2_{\rm in} }}
-{{\partial^2}\over{\partial \theta^2
}}-2i\csc\theta{{\partial^2}\over{\partial \theta \partial
\varphi_{\rm in} }}
\nonumber \\
&+&{{2i\cos\theta}\over{\sin ^2 \theta }}{{\partial}\over{\partial
\varphi_{\rm in} }}+\cot \theta {{\partial}\over{\partial \theta
}}+2ia \sin \theta{{6r-2ia \cos \theta} \over{\Sigma}}[ia \sin
\theta {{\partial}\over{\partial t_{\rm in}}}
+{{\partial}\over{\partial \theta}}+ i \csc \theta
{{\partial}\over{\partial \varphi_{\rm
in}}} \nonumber \\
&-&ia\sin\theta\rho^*_{\rm in}]
\rbrace,  \nonumber \\
{}_{2}{\cal{T}}^{mm}&=& -\lbrace ({{\partial}\over{\partial t_{\rm in}
}}-{{\partial}\over{\partial r }} -{{6r-10ia \cos \theta} \over
{\Sigma}})({{\partial}\over{\partial t_{\rm in} }}-{{\partial}\over{\partial r }})
+{{4}\over{\Sigma^2}}(r^2-5 a^2 \cos ^2 -4ria\cos\theta)
\rbrace,\nonumber \\
{}_{2}{\cal{T}}^{lm}&=& {{-2\rho_{\rm in}}\over{\sqrt{2}}}\lbrace
(ia\sin \theta{{\partial}\over{\partial t_{\rm in} }}
+{{\partial}\over{\partial\theta }} +i\csc \theta
{{\partial}\over{\partial \varphi_{\rm in} }}+\cot\theta +(2r-3ia
\cos \theta){{ia \sin \theta} \over
{\Sigma}})({{\partial}\over{\partial t_{\rm in}
}}-{{\partial}\over{\partial r}} )\nonumber \\
&-&3\rho_{\rm in}(ia \sin \theta {{\partial}\over{\partial t_{\rm
in} }}+{{\partial}\over{\partial \theta }} +i \csc \theta
{{\partial}\over{\partial \varphi_{\rm in} }}-\cot \theta)+{{ia
\sin \theta \rho_{\rm in}} \over {\Sigma}}
(6r^2+5iar \cos \theta - 5a^2\cos^2 \theta) \rbrace, \nonumber \\
{}_{-2}{\cal{T}}^{nn}&=& {{\rho^{*2}_{\rm in}}\over{2}}\lbrace
a^2\sin^2\theta{{\partial^2}\over{\partial t^2_{\rm in} }}
+2ia\sin\theta{{\partial^2}\over{\partial\theta \partial t_{\rm
in} }} +2a{{\partial^2}\over{\partial \varphi_{\rm in} \partial
t_{\rm in} }} +\csc^2\theta{{\partial^2}\over{\partial
\varphi^2_{\rm in} }} -{{\partial^2}\over{\partial \theta^2 }}
+2i\csc\theta{{\partial^2}\over{\partial \theta \partial
\varphi_{\rm in} }} \nonumber \\
&-&\cot \theta {{\partial}\over{\partial \theta}}  +2ia \cos
\theta {{\partial}\over{\partial t}} +ia{\rho^{*}_{\rm in}\sin
\theta}[ia \sin \theta {{\partial}\over{\partial t_{\rm in}}}
-{{\partial}\over{\partial \theta_{\rm in}}} +i \csc \theta
{{\partial}\over{\partial \varphi_{\rm in}}}]
-2a^2 \sin^2 \theta \rho^{2}_{\rm in} \rbrace \, \rho^{-4}_{\rm in},  \nonumber \\
{}_{-2}{\cal{T}}^{m^*m^*}&=&-{{1} \over {4\Sigma^2}}\lbrace
\Delta_+(\Delta_+{{\partial}\over{\partial t_{\rm in} }} +2\Delta
{{\partial}\over {\partial r }} +4a{{\partial}\over{\partial
\varphi_{\rm in}}}){{\partial}\over{\partial t_{\rm in} }}
+\Delta(\Delta{{\partial}\over {\partial r }}+4a
{{\partial}\over{\partial
\varphi_{\rm in}}}){{\partial}\over{\partial r }}  +
4a^2{{\partial^2}\over{\partial \varphi^2_{\rm in} }}\nonumber \\
&+&4M(3r^2+a^2){{\partial}\over{\partial t_{\rm in}
}}-2(r-3ia\cos\theta){{\Delta}\over
{\Sigma}}(\Delta_+{{\partial}\over{\partial t_{\rm in} }} +\Delta
{{\partial}\over {\partial r }} +2a{{\partial}\over{\partial
\varphi_{\rm in}}})-8ia\cos\theta\rho_{\rm
in}{{\Delta^2}\over{\Sigma}}
\rbrace\,\rho^{-4}_{\rm in}, \nonumber \\
{}_{-2}{\cal{T}}^{nm^*}&=&  {{ \rho^*_{\rm in}} \over { \sqrt{2}
\Sigma}}\lbrace (\Delta_+{{\partial}\over{\partial t_{\rm in} }} +
\Delta {{\partial}\over{\partial r}} +2a{{\partial}\over{\partial
\varphi_{\rm in}}} +
(2r+3ia\cos\theta){{\Delta}\over{\Sigma}}-r+M)( ia \sin \theta
{{\partial}\over{\partial t_{\rm in} }} -{{\partial}\over{\partial
\theta }}\nonumber\\
&+&i \csc \theta {{\partial}\over{\partial \varphi_{\rm in }}}) +(
\cot\theta -
(3r+ia\cos\theta){{ia\sin\theta}\over{\Sigma}})(\Delta_+{{\partial}\over{\partial
t_{\rm in} }} + \Delta {{\partial}\over{\partial r}}
+2a{{\partial}\over{\partial \varphi_{\rm in}}})
+6iar\sin\theta\rho_{\rm in}{{\Delta}\over{\Sigma^2}}
\nonumber\\
&-&2(r-M)( \cot\theta
-(3r+ia\cos\theta){{ia\sin\theta}\over{\Sigma}})-
(r+3ia\cos\theta)\cot\theta{{\Delta}\over{\Sigma}}\rbrace
\rho^{-4}_{\rm in},   \label{eq:ffin}
\eea 
\noindent where we have also act the ${}_{-2}{\cal{T}}^{\mu\nu}$ operator on $\rho^4$.

Once obtained these perturbation equations for arbitrary sources within a
description free of geometric divergencies, it is possible to
proceed further and perform the numerical evolution of the ${\Psi_0}^{(1)}$ and ${\Psi_4}^{(1)}$ perturbed
Weyl scalars for cases as interesting as the ones due to an orbiting object, or the
one mention above of an inspiring and finally colliding and coalescing object, or
those due to accreting matter, for mention some examples. These studies are currently
under way.

We finish this presentation testing the perturbation equations Eqs.(\ref{eq:Tfin}) in
Schwarzschild described with the Kerr Schild coordinates, which implies setting the
angular parameter $a=0$, and the Eqs.(\ref{eq:Tfin},\ref{eq:ffin}) reduce to the form:

\bea &&\lbrack \Delta_+{{\partial^2}\over {\partial t^2_{\rm
in}}}- 4\,M\,r\,{{\partial^2}\over{\partial r \partial t_{\rm in}
}} -\Delta\,{{\partial^2}\over{\partial r^2}}
-{{\partial^2}\over{\partial
\theta^2}}-{{1}\over{\sin^2\theta}}{{\partial^2}\over{\partial
\varphi^2_{\rm in}}} -
2\left((s+1)(r-M)+2s\Delta\,\varepsilon_{\rm in}\right){{\partial}\over{\partial r}} \nonumber \\
&& -2\left(s\,r+(s+1)M\right){{\partial}\over{\partial t_{\rm
in}}}-\cot\theta\,{{\partial}\over{\partial \theta}}
-2si{{\cos\theta}\over{\sin^2\theta}}{{\partial}\over{\partial
\varphi_{\rm in}}}
-s(1-s\cot^2\theta)\rbrack{}_s\psi=16\pi\,\Sigma\,{}_{s}{\cal{T}}{^{\mu\nu}}\,{T_{\mu\nu}}^{(1)},
\nonumber \\ \label{eq:Tfins} \eea with $\Delta_+=r^2+2\,M\,r$,
the angular coordinate $\varphi$ remains unchanged in this case,
and now $\Delta=r^2-2\,M\,r$. The operators
${}_{s}{\cal{T}}{^{\mu\nu}}$ are given by Eq.(\ref{eq:tau1}) but
now the projected operators reduce to:

\bea {}_{2}{\cal{T}}^{ll}&=&
 {1\over{2r^2}}\lbrace
\csc^2\theta{{\partial^2}\over{\partial \varphi^2_{\rm in} }}
-{{\partial^2}\over{\partial \theta^2
}}-2i\csc\theta{{\partial^2}\over{\partial \theta \partial
\varphi_{\rm in}}}+{{2i\cos\theta}\over{\sin ^2 \theta
}}{{\partial}\over{\partial \varphi_{\rm in}}}
+\cot \theta {{\partial}\over{\partial \theta }}\rbrace,  \nonumber \\
{}_{2}{\cal{T}}^{mm}&=& -\lbrace ({{\partial}\over{\partial t_{\rm
in} }}-{{\partial}\over{\partial r }} -{{6}\over
{r}})({{\partial}\over{\partial t_{\rm in}
}}-{{\partial}\over{\partial r }}) -{{4}\over{r^2}}
\rbrace,\nonumber \\
{}_{2}{\cal{T}}^{lm}&=& -{{2}\over{\sqrt{2}r}}\lbrace
 ({{\partial}\over{\partial t_{\rm in}}}-{{\partial}\over{\partial r}}-
 {3\over{r}})({{\partial}\over{\partial \theta}}+
i\csc\theta {{\partial}\over{\partial \varphi_{\rm in}}}+\cot\theta)  \rbrace, \nonumber \\
{}_{-2}{\cal{T}}^{nn}&=& {{1}\over{2r^2}}\lbrace
\csc^2\theta{{\partial^2}\over{\partial \varphi^2_{\rm in}}}
-{{\partial^2}\over{\partial \theta^2 }}
+2i\csc\theta{{\partial^2}\over{\partial \theta \partial
\varphi_{\rm in} }}
+\cot \theta {{\partial}\over{\partial \theta}}  \rbrace \, r^{4},  \nonumber \\
{}_{-2}{\cal{T}}^{m^*m^*}&=&-{{1} \over {4r^4}}\lbrace
\Delta_+(\Delta_+{{\partial}\over{\partial t_{\rm in} }} +2\Delta
{{\partial}\over {\partial r }}){{\partial}\over{\partial t_{\rm
in} }} +\Delta^2{{\partial^2}\over {\partial r^2 }}
+12Mr^2{{\partial}\over{\partial t_{\rm in} }}-2{{\Delta}\over
{r}}(\Delta_+{{\partial}\over{\partial t_{\rm in} }} +\Delta
{{\partial}\over {\partial r}})
\rbrace\,r^4, \nonumber \\
{}_{-2}{\cal{T}}^{nm^*}&=&  {{1} \over{ \sqrt{2}r^3}}\lbrace
\left(\Delta_+{{\partial}\over{\partial t_{\rm in} }} + \Delta
{{\partial}\over{\partial r}}  +
2{{\Delta}\over{r}}-r+M\right)\left( -{{\partial}\over{\partial
\theta }}+i \csc \theta {{\partial}\over{\partial \varphi_{\rm in
}}}\right)\nonumber\\
&&+ \cot\theta \left(\Delta_+{{\partial}\over{\partial t_{\rm in}
}} + \Delta {{\partial}\over{\partial r}}\right)
-2(r-M)\cot\theta- \cot\theta{{\Delta}\over{r}}\rbrace r^{4}.
\label{eq:ffinsch}
\eea

We try the simple case when the perturbation is due to dust
radially falling unto the black hole, for which the stress energy
tensor is given by \be T^{\mu\nu}=d(r,t)\,u^\mu\,u^\nu,
\label{eq:ste} \ee with $d(r,t)$ the density and $u^\mu$ the four
velocity, which for the radially falling case has only radial and
temporal components: \be u^\mu=(u^t,u^r,0,0).\label{eq:u4} \ee
Projecting the stress energy tensor, Eq.(\ref{eq:ste}), unto the
null tetrad which for the Schawrzschild case is given by: (from
Eq.(\ref{eq:nulltin}) with $a=0$)
\begin{eqnarray}
l_{\mu}&=&\left( 1,-1,0,0 \right), \nonumber \\
n_{\mu}&=&\frac{1}{2 r^2}\left(\Delta,-(r^2+2Mr),0,0 \right), \nonumber \\
m_{\mu}&=&\frac{r}{\sqrt{2}}\left(0,0,1, i\,\sin\theta \right),
\end{eqnarray}
and taking into account the components different from zero of the
four velocity, Eq.(\ref{eq:u4}), we obtain that the stress energy
tensor projections unto the null tetrad are, \bea
T_{ll}&=&d^2(r,t)\,(u^0-u^r)^2, \nonumber \\
T_{nn}&=&{{d^2(r,t)}\over{4r^4}}(\Delta\,u^0-\Delta_+\,u^r)^2, \nonumber \\
T_{lm}&=&T_{mm}=T_{nm^*}=T_{m^*m^*}=0. \eea
 Notice that the non zero projections are
functions of $r$ and $t$ only. Thus, substituting the projections in the rhs of
Eqs.(\ref{eq:Tfins}), there is only left the action of the operators on $T_{ll}$ and
on $T_{nn}$. But those operators for the Schwarzschild case are purely angular
operators, and as the projections are functions of $(t,r)$, the action is zero. In
this way, we conclude that radially infalling dust into the Schwarzschild black hole
does not produce perturbations on the ${\Psi_0}^{(1)}$ and ${\Psi_4}^{(1)}$ perturbed
Weyl scalars for both cases.

This result is expected considering that the gravitational waves carry quadrupolar radiation and higher,
and that the radially infalling dust into the Schwarzschild black hole will produce only monopolar
perturbations, which will not be reflected in the ${\Psi_0}^{(1)}$ and ${\Psi_4}^{(1)}$ perturbed
Weyl scalars.

It can be seen that the same happens with a radially infalling massive scalar field, with stress energy
tensor given by
\be
T_{\mu\nu}=\phi_{\mu}\,\phi_{\nu}-{1\over{2}}g_{\mu\nu}\left( \phi^\alpha\phi_\alpha-2V(\phi)\right),
\ee
with the scalar field a function of $(t,r)$ only, and where subindex and superindex imply derivative
with respect to $x^\mu$. Again, the only non null projections needed in the perturbation equation are
\begin{equation}
T_{ll}=(\dot{\phi}+\phi^\prime)^2 \qquad {\rm and} \qquad
T_{nn}={{1}\over{4r^4}}(\Delta_+
\,\dot{\phi}-\Delta\,\phi^\prime)^2,
\end{equation}
where dot and prime stand for derivatives with respect to $t$ and $r$ respectively. Once more,
they are only functions of $(t,r)$, thus the action of the operators on them is zero, so the
massive scalar field radially infalling unto the Schwarzschild black hole does not produce
perturbations on the ${\Psi_0}^{(1)}$ and ${\Psi_4}^{(1)}$ perturbed Weyl scalars.

\subsection{Outgoing case}

In this section we present the analogous derivation of the field equations for the
perturbed scalar projections ${\Psi_0}^{(1)}$ and ${\Psi_4}^{(1)}$  of the Weyl
tensor, including sources. The procedure, although length, is straightforward and
makes the corresponding steps as in the ingoing case. Thus, from the null vector
naturally defined in the Kerr Schild form, Eq.(\ref{eq:l}), we construct the following
contravariant null tetrad:
\begin{eqnarray}
{l_{\mu}}_{\rm out} &=&\left( 1,-1,0,-a\sin^{2}\theta \right), \nonumber \\
{n_{\mu}}_{\rm out} &=&\frac{1}{2 \Sigma}\left(\Delta,-(r^2+a^2(\cos^2\theta-\sin^2\theta)+2Mr),0,-
a\Delta\sin^2\theta \right), \nonumber \\
{m_{\mu}}_{\rm
out}&=&\frac{1}{\sqrt{2}\,(r+i\,a\,\cos\theta)}\left[-i\,a\,\sin\theta,i\,a\,\sin\theta,\Sigma,
i\,(r^2+a^2)\,\sin\theta \right].
\end{eqnarray}

Thus, the covariant null tetrad is:
\begin{eqnarray}
{l^{\mu}}_{\rm out} &=&\left( 1,1,0,0 \right), \nonumber \\
{n^{\mu}}_{\rm out} &=&\frac{1}{2 \Sigma}\left( r^2+a^2+2Mr, -\Delta ,0,2a \right) ,\nonumber  \\
{m^{\mu}}_{\rm out} &=&-\frac{1}{\sqrt{2}\,(r+i\,a\,\cos\theta)}(ia\sin
\theta,0,1,i\csc \theta).
\end{eqnarray}

From which we obtain the corresponding differential operators:
\begin{eqnarray}
\bf{D}_{\rm out} &=&{l^{\mu}}_{\rm out}\partial _{\mu}=\partial_{t_{\rm out}} +\partial_r,\nonumber \\
\bf{\Delta}_{\rm out} &=&{n^{\mu}}_{\rm out}\partial_{\mu}=\frac{1}{2
\Sigma}\left[(r^2+a^2+2Mr)\partial_{t_{\rm out}}-\Delta\partial_r+
2a\partial_{\varphi_{\rm out}} \right], \nonumber\\
\bf{\delta}_{\rm out}  &=&{m^{\mu}}_{\rm
out}\partial_{\mu}=\frac{{\rho^*}_{\rm out}}{\sqrt{2}}
(ia\sin \theta\partial_{t_{\rm out}}+\partial_\theta+i\csc
\theta\partial_{\varphi_{\rm out}}), \label{op:out}
\end{eqnarray}
with the spinor coefficient $\rho_{out}$ defined below in Eq.(\ref{eq:espout}).

The spinor coefficients are computed directly from their definition
\cite{libro,frol}, obtaining:
\begin{eqnarray}
&&\kappa_{\rm out}=\sigma_{\rm out}=\lambda_{\rm out}=\nu_{\rm out}=\varepsilon_{\rm out}=0,  \nonumber \\
&&\rho_{\rm out}=-\frac{1}{r-i\,a\,\cos\theta},\quad \mu_{\rm
out}=\frac{\Delta}{2\Sigma}\,\rho_{\rm out},
\quad \gamma_{\rm out}=\mu +\frac{r-M}{2\,\Sigma}, \nonumber \\
&&\pi_{\rm out} =-\frac{ia\sin \theta}{\sqrt{2}}\,{\rho_{\rm
out}}^2, \quad \beta_{\rm out}=\frac{\cot
\theta}{2\sqrt{2}}\,{\rho_{\rm out}}^*, \quad \alpha_{\rm out}
=\pi_{\rm out}-{\beta^*}_{\rm out}, \quad \tau_{\rm out}
=\frac{ia\sin \theta}{\sqrt{2}\,\Sigma}, \label{eq:espout}
\end{eqnarray}
and finally, the only non zero Weyl scalar is:
\begin{equation}
{\Psi _{2}}_{\rm out}=M{\rho_{{\rm out}}}^3. \label{eq:yaya}
\end{equation}

As mentioned above, the Teukolsky equation, that is the field equation the perturbed
scalar projections ${\Psi_0}^{(1)}$ and ${\Psi_4}^{(1)}$  of the Weyl tensor,
including sources, Eq.(\ref{eq:Tf}), as well as the expression for the operators
action on the sources, Eq.(\ref{eq:tau1}), are valid also for the outgoing case.
Thus, using the geometric quantities given by
Eqs.(\ref{op:out},\ref{eq:espout},\ref{eq:yaya}), and their properties given in the
appendix B, Eq.(\ref{a:spin},\ref{a:spinor}) after a quite lengthily algebraic manipulation, we obtain the following
equation for ${\Psi_0}^{(1)}$ and ${\Psi_4}^{(1)}$ perturbed Weyl scalars due to
arbitrary source terms in outgoing coordinates:

\bea &&\lbrack (\Sigma+2\,M\,r){{\partial^2}\over {\partial
t^2_{\rm out}}}+ 4\,M\,r\,{{\partial^2}\over{\partial r \partial
t_{\rm out} }} -\Delta\,{{\partial^2}\over{\partial
r^2}}+2\,a\,{{\partial^2}\over{\partial r\partial \varphi_{\rm
out} }} -{{\partial^2}\over{\partial
\theta^2}}-{{1}\over{\sin^2\theta}}{{\partial^2}\over{\partial
\varphi^2_{\rm out}}} \nonumber \\
&&- 2(s+1)(r-M){{\partial}\over{\partial r}} +2\left(
s\,r+(s+1)M+s\,ia\cos\theta \right){{\partial}\over{\partial
t_{\rm out }}}-\cot\theta{{\partial}\over{\partial \theta}}
\nonumber \\
&&-2(s+1)i{{\cos\theta}\over{\sin^2\theta}}{{\partial}\over{\partial
\varphi_{\rm out}}}
-s(1-s\cot^2\theta)\rbrack{}_s\psi=16\pi\,\Sigma\,{}_{s}{\cal{T}}{^{\mu\nu}}\,{T_{\mu\nu}}^{(1)},
\label{eq:Tfino} \eea

\noindent where ${}_{s}{\cal{T}}{^{\mu\nu}}$ are given again by
Eq.(\ref{eq:tau1}), but using the outgoing terms, the projected
operators have now the following expressions: \bea
{}_{2}{\cal{T}}^{ll}&=&-{{{{\rho^*}^2}_{\rm out}}\over{2}}\lbrack
ia\sin\theta(ia\sin\theta{{\partial}\over {\partial t_{\rm
out}}}+2{{\partial}\over {\partial
\theta}}+2i\csc\theta{{\partial}\over {\partial \varphi_{\rm
out}}}){{\partial}\over {\partial t_{\rm
out}}}+\sin\theta{{\partial}\over {\partial
\theta}}(\csc\theta{{\partial}\over {\partial
\theta}}) \nonumber \\
&&+2i{{\partial}\over {\partial
\theta}}(\csc\theta{{\partial}\over {\partial \varphi_{\rm
out}}})-\csc^2\theta{{\partial^2}\over{\partial \varphi^2_{\rm
out}}}+2{{ia\sin\theta}\over{\Sigma}}(3r+ia\cos\theta)(ia\sin\theta{{\partial}\over
{\partial t_{\rm out}}} \nonumber \\ &&+{{\partial}\over {\partial
\theta}}+i\csc\theta{{\partial}\over {\partial \varphi_{\rm
out}}}-ia\sin\theta\,\rho^*_{\rm out}) \rbrack,\nonumber \\
{}_{2}{\cal{T}}^{mm}&=&- ({{\partial}\over {\partial t_{\rm
out}}}+{{\partial}\over {\partial
r}}+{{3r+ia\cos\theta}\over{\Sigma}})({{\partial}\over {\partial
t_{\rm
out}}}+{{\partial}\over {\partial r}})-{4\over{\Sigma}},\nonumber \\
{}_{2}{\cal{T}}^{lm}&=&-{{{\rho^*}_{\rm
out}}\over{\sqrt{2}}}\lbrack ({{\partial}\over {\partial t_{\rm
out}}}+{{\partial}\over {\partial
r}}+{{3r+ia\cos\theta}\over{\Sigma}})(ia\sin\theta{{\partial}\over
{\partial t_{\rm out}}}+{{\partial}\over {\partial
\theta}}+i\csc\theta{{\partial}\over {\partial \varphi_{\rm
out}}}-\cot\theta)\nonumber \\ &&-2ia\sin\theta(2\rho_{\rm
out}{{\partial}\over {\partial t_{\rm
out}}}+{{\rho^*}^2}_{\rm out}) \rbrack, \nonumber \\
{}_{-2}{\cal{T}}^{nn}&=&-{{{\rho^2}_{\rm out}}\over{2}}\lbrack
ia\sin\theta(ia\sin\theta{{\partial}\over {\partial t_{\rm
out}}}-2{{\partial}\over {\partial
\theta}}+2i\csc\theta{{\partial}\over {\partial \varphi_{\rm
out}}}){{\partial}\over {\partial t_{\rm
out}}}+\sin\theta{{\partial}\over {\partial
\theta}}(\csc\theta{{\partial}\over {\partial
\theta}}) \nonumber \\
&& -\csc^2\theta{{\partial^2}\over{\partial \varphi^2_{\rm
out}}}-{{ia\sin\theta}\over{\Sigma}}(7r+11ia\cos\theta)(ia\sin\theta{{\partial}\over
{\partial t_{\rm out}}}-{{\partial}\over {\partial
\theta}}+i\csc\theta{{\partial}\over {\partial \varphi_{\rm
out}}})\nonumber \\
&&-2i{{\partial}\over {\partial
\theta}}(\csc\theta{{\partial}\over {\partial \varphi_{\rm
out}}})+2\rho_{\rm out}{{a^2\sin^2\theta}\over{\Sigma}}(r+9ia\cos\theta) \rbrack ,  \nonumber \\
{}_{-2}{\cal{T}}^{m^*m^*}&=&-{1\over{\Sigma^2}}\lbrack
\Delta_+(\Delta_+{{\partial}\over {\partial t_{\rm out}}}
-2\Delta{{\partial}\over {\partial r}} +4a{{\partial}\over
{\partial \varphi_{\rm out}}}){{\partial}\over {\partial t_{\rm
out}}}+ \Delta(\Delta{{\partial}\over {\partial r}}
-4a{{\partial}\over {\partial \varphi_{\rm out}}}){{\partial}\over
{\partial r}} +4a^2{{\partial^2}\over {\partial \varphi^2_{\rm
out}}}
\nonumber \\
&&+4M(r^2-a^2){{\partial}\over {\partial t_{\rm out}}}
-2{{\Delta}\over{\Sigma}}(3r+5ia\cos\theta)
(\Delta_+{{\partial}\over {\partial t_{\rm out}}}
-\Delta{{\partial}\over {\partial r}} +2a{{\partial}\over
{\partial \varphi_{\rm out}}}-2ia\cos\theta{{\Delta}\over{\Sigma}})\nonumber \\
&&+4a(r-M){{\partial}\over {\partial \varphi_{\rm out}}}+4r(r+ia\cos\theta)({{\Delta}\over{\Sigma}})^2\rbrack, \nonumber \\
{}_{-2}{\cal{T}}^{nm^*}&=& -{{{\rho^*}_{\rm
out}}\over{2\sqrt{2}\Sigma}}\lbrack (\Delta_+{{\partial}\over
{\partial t_{\rm out}}} -\Delta{{\partial}\over {\partial r}}
+2a{{\partial}\over {\partial \varphi_{\rm
out}}}+2(r-M)-(5r+3ia\cos\theta){{\Delta}\over{\Sigma}} )\times
\nonumber
\\
&& \times (ia\sin\theta{{\partial}\over {\partial t_{\rm out}}}-{{\partial}\over
{\partial \theta}}+i\csc\theta{{\partial}\over {\partial \varphi_{\rm
out}}}+\cot\theta -(3r+5ia\cos\theta{{ia\cos\theta}\over{\Sigma}})) \nonumber
\\
&&
-3ia\sin\theta(r^2+6ia\cos\theta-a^2\cos^2\theta){{\Delta}\over{\Sigma^2}}\rbrack.
\label{eq:ffout} \eea

Again we recall that these equations were directly obtained working from the original
field perturbation equation, Eqs.(\ref{eq:Tfin}).

\section{Conclusions}

Scratching form the top, we have derived the equations for two of the radiative parts
of the perturbed Weyl tensor, namely ${\Psi_0}^{(1)}$ and ${\Psi_4}^{(1)}$ in the
Kerr spacetime described in the Kerr Schild coordinates, including sources. We have
done son for the ingoing as well as for the outgoing descriptions.

Even thought the procedure followed in this work is straight forward, we consider that
the equations obtained are a needed step in the research line of semi analytical
studies of perturbed black holes, and will be useful for community.

The present description allows us to perform studies as near to the horizon as is
wished, even inside the horizon for that matter, as long as the coordinates are
regular. Also, there are several works which describe initial data, namely in the Kerr
Schild description, so our work fits well with them. We remark the fact that the
derived equations with sources presented in this work were obtained directly from the
Bianchi identities within the general null tetrad formulation.  The reason for this
is that the spinors operators are not scalar quantities and the changes from one
tetrad to another are not due to coordinate transformations only. Thus, it is not
safe to use the equations for the perturbed components derived by S. Teukolsky, back
in 1972 \cite{Teukolsky}, where he used the Eddington Finkelstein coordinates (a
description well suited for the asymptotic analysis, which is the one he performed),
and use the coordinate transformation to describe those equation in the Kerr Schild
coordinates.

Thus, we derived the equations and tested them in simple cases where we obtain the
expected results, such as the fact that radial perturbations do not generate
radiation. The development of realistic situations is currently under way. This work
continues the program started by Campanelli et al. \cite{Pablo2} where they studied
the gravitational perturbation within a description free of coordinate singularities,
without sources. We have introduce another null tetrad and derived the equations with
arbitrary sources.

The analysis and the equations presented in this work are a complement for those
performed purely numerically. Joint efforts are currently under way. As mentioned
above, there are several phenomena that can now be studied from this perspective,
mainly the most important is to continue in the study of gravitational waves from a
compact star in circular, inspiral orbits around a massive spinning black hole, with
applications to observation by LISA and LIGO \cite{Sam}, that is, to obtain
predictions which could be detected and tested by the gravitational observatories
which will start to work in the next year.

\section{Acknowledgments}

This work was partially supported by a CONACyT scholarship, and the grant
DGAPA-UNAM IN121298. We want to thank Pablo Laguna for fruitful discussion during the elaboration of the present work.

\section{Appendix A}

The following commutation relations between the operators are
repeatedly used in the work, and are valid for any algebraically
special type D spacetime (we used that $\alpha=\pi-\beta^*$): \bea
&{\bf D}\,{\bf \Delta}={\bf \Delta}{\bf D}-(\gamma+\gamma^*)\,{\bf
D}-(\varepsilon+\varepsilon^*)\,{\bf \Delta}
+(\tau+\pi^*)\,{\bf \delta}+(\tau^*+\pi)\,{\bf \delta^*},& \nonumber \\
&{\bf \delta}\,{\bf \delta}^*={\bf \delta}^*{\bf \delta}+(\mu-\mu^*)\,{\bf
D}+(\rho-\rho^*)\,{\bf \Delta}
+(2\,\beta-\pi^*)\,{\bf \delta}-(2\,\beta^*-\pi)\,{\bf \delta^*},& \nonumber \\
&\left({\bf D}+q\,\rho-\rho^*-(p+1)\varepsilon+ \varepsilon^*\right)\, \left({\bf
\delta}-p\beta+q\tau \right)-
\left({\bf \delta}-p\beta+q\tau \right)\,\left({\bf D}+q\,\rho-p\varepsilon \right)=0,& \nonumber \\
&\left({\bf \Delta}-q\,\mu+\mu^*+(p+1)\gamma - \gamma^*\right)\,
\left({\bf \delta}^*-p\beta^*+(p-q)\pi \right) \nonumber \\ &
-\left({\bf \delta}^*-p\beta^*+(p-q+1)\pi-\tau^*
\right)\,\left({\bf \Delta}-q\mu+p\gamma \right)=0,& \label{a:con}
\eea where $p$ and $q$ are arbitrary real numbers. This expression
are valid for the outgoing and ingoing case.

Finally, the following identities among the spinor coefficients are also useful: \bea
\begin{array}{llll}
\beta^*\,\tau=-\beta\,\pi,&& \pi^*\,\rho=\tau\,\rho^*, \nonumber \\
\mu^*\,\rho=\mu\,\rho^*, &&
2\,\rho\,\gamma-4\,\beta\,\pi-2\,\pi\,\tau-{\Psi_2}=0. 
\end{array} \\
\eea

\section{Appendix B}

In the derivation of the field equations, is useful to determine the action of the
operators on the different spinor coefficients. Its explicit form is obtained from
teh expressions for the spinor coefficients given by Eq.(\ref{eq:espin},
\ref{eq:espout}), and the operators given by Eq.(\ref{op:in}, \ref{op:out}). Some of
those actions are valid for both cases, the ingoing and the outgoing:

\bea
\begin{array}{lllll} {\bf D}\rho^*=\rho^{*2}, && {\bf
\Delta}\tau^*=-(\mu+\mu^*)\,\tau^*, && {\bf
\delta}\rho=\tau\,\rho, \nonumber \\
{\bf \delta}^*\rho=-\pi\,\rho, && {\bf
\delta}\pi^*=2(\beta-\pi^*)\,\pi^*, && {\bf
\delta}^*\mu=-(2\pi-\tau^*)\,\mu\, , \nonumber \\
{\bf \delta}\rho^*=\pi\,\rho,&& {\bf \delta}^*\pi=2(\beta^*-\pi)\,\pi, && {\bf
\delta}^*\tau^*=(2\beta^*-\pi+\tau^*)\,\tau^*. \label{a:spin}
\end{array} \\
\eea

\noindent The following expressions are valid the ingoing case:
\bea
\begin{array}{lllll}
{\bf D}\varepsilon={{1}\over{2}}(\rho^2-\rho^{-2}), && {\bf
D}\beta={{\rho}\over{2}}(2\beta+\tau-\tau^*), &&
{\bf D}\pi^*=\rho(\tau^*-\pi), \nonumber \\
{\bf \Delta}\tau^*=-2\mu^*\tau^*,
&& {\bf \Delta}\pi=\pi^*(\mu+\mu^*), && {\bf \Delta}\mu=-(\mu+2\gamma)\,\mu, \nonumber \\
{\bf \Delta}\mu^*=-(\mu^*+2\gamma)\,\mu^*, && {\bf
\Delta}\beta=-{{\mu^*}\over{2}}(2\beta^*+\tau-\tau^*),
&& {\bf \delta}\mu=\mu(2\tau-\pi^*), \nonumber \\
{\bf \delta}\varepsilon={{\rho}\over{2}}(\tau-\tau^*), && {\bf
\delta}^*\mu^*=(2\tau^*-\pi)\,\mu^*\, ,  && {\bf
\delta}^*\Sigma^{-1}=2\rho \tau^* \epsilon^*.\label{a:spinII}
\end{array} \\
\eea

\noindent And, for the outgoing case, the following expressions
are valid \bea
\begin{array}{lllll}
{\bf D}\beta=\beta\,\rho^*, && {\bf D}\pi^*=2\,\pi^*\rho^*, &&
{\bf \Delta}\tau^*=-(\mu+\mu^*)\,\tau^*, \nonumber \\
{\bf \Delta}\pi=-2\,\pi\,\mu, && {\bf \Delta}\mu=-(\mu^*+2\gamma)\,\mu,
&& {\bf \Delta}\mu^*=-(\mu+2\gamma)\,\mu^*, \nonumber \\
{\bf \Delta}\beta=-\mu^*\beta^*, && {\bf
\delta}\mu=\mu^*(2\pi^*-\tau), && {\bf
\delta}^*\mu^*=-(2\pi^*+\tau)\,\mu\, , \nonumber \\
{\bf \delta}\beta^*=-\mu\,\beta^*.  && && \label{a:spinor}
\end{array} \\
\eea

\end{document}